\documentclass{article}

\usepackage{graphicx}
\usepackage{amsmath}
\usepackage{authblk}
\usepackage{natbib}
\usepackage{caption}

\title{Wetting of Water on Graphene}
\author[1]{Bijoyendra Bera}
\author[1]{Noushine Shahidzadeh}
\author[2,3,4]{Himanshu Mishra}
\author[1]{Daniel Bonn}
\affil[1]{Soft Matter Group, Institute of Physics, Science Park 904\\
 1098XH Amsterdam, The Netherlands}
\affil[2]{King Abdullah University of Science and Technology (KAUST)\\
$^3$Water Desalination and Reuse Center (WDRC)\\
$^4$Biological and Environmental Sciences \& Engineering Division (BESE), Thuwal, 23955-6900, Saudi Arabia}
\date{}

\begin{document}

\maketitle

\section*{Abstract}

The wetting properties of graphene have proven controversial and difficult to assess. The presence of a graphene layer on top of a substrate does not significantly change the wetting properties of the solid substrate, suggesting that a single graphene layer does not affect the adhesion between the wetting phase and the substrate. However, wetting experiments of water on graphene show contact angles that imply a large amount of adhesion. Here, we investigate the wetting of graphene by measuring the mass of water vapor adsorbing to graphene flakes of different thickness at different relative humidities. Our experiments unambiguously show that the thinnest of graphene flakes do not adsorb water, from which it follows that the contact angle of water on these flakes is $\sim$180$^o$. Thicker flakes of graphene nanopowder, on the other hand, do adsorb water. A calculation of the van der Waals (vdW) interactions that dominate the adsorption in this system confirms that the adhesive interactions between a single atomic layer of graphene and water are so weak that graphene is superhydrophobic. The observations are confirmed in an independent experiment on graphene-coated water droplets that shows that it is impossible to make liquid `marbles' with molecularly thin graphene.

\newpage

\section*{Introduction}

Atomically thin layers of graphene ($sp^2$ hybridized carbon atoms in a hexagonal honeycomb array) exhibit an exceptionally rich range of mechanical and electronic properties such as high mechanical strength, impenetrability, chemical stability, optical transparency, and high electrical and thermal conductivity \cite{gomez,shih2}. As a result, graphene has become one of the most active areas of scientific research in the last decade \cite{novoselov, geim1, geim2}. Among many others, potential applications of graphene in water desalination \cite{castroneto}, electrochemistry \cite{dresselhaus} and catalysis \cite{oshima} are being explored intensively. However, a clear understanding of the physical interactions between graphene and water has remained elusive. Recent experiments have shown that the contact angle of water on graphene-coated substrates and bare substrates are almost identical, which suggests that graphene is wetting-transparent \cite{rafiee}. Substrate interactions play an important role in these experiments. To avoid these, and assess the wetting properties of water on graphene only, we use an entirely different experimental approach, involving only graphene and no substrates. We measure the water adsorption from water vapor on graphene nanopowders consisting of nanoflakes of different thicknesses. For molecularly thin graphene, we find that that the contact angle of water is indistinguishable from 180$^o$; we show that this is indeed due to the adhesive interactions from a molecular layer being vanishingly small.\\
The `wetting transparency' of graphene was first reported by Rafiee \textit{et al.} \cite{rafiee} based on static contact angle measurements with water drops on graphene-coated $Si$ and $Au$ substrates. On $Si$ and $Au$, the contact angle of a water drop is shown to be $\sim$33$^o$ and $\sim$77$^o$, respectively, both with and without a graphene layer present on the substrate. They argued that the atomically thin layer of graphene is not relevant for the adhesive van der Waals interaction. Subsequently, a debate ensued regarding the nature of the interactions. Calculations of the van der Waals and short range repulsive forces suggest that a graphene layer would alter the wetting properties on superhydrophilic and a superhydrophobic substrates \cite{shih}. Contact angle measurements by Raj \textit{et al.}, \cite{raj} on the other hand, demonstrated that contact angles are independent of the number of graphene layers on a glass or silicon oxide substrate. They also pointed out the influence of the substrate material in estimating the wettability of graphene coated substrates. The possible role of contamination has also been extensively discussed \cite{li,xu}.\\

In all the experiments and simulations mentioned above, the wetting phase water is invariably introduced in the form of a sessile drop \cite{checco}. However, there is another way of assessing the wetting properties of a liquid on a given substrate \cite{bonn1}; in this work, we measure the thickness of adsorbed water films from the vapor phase as the relative humidity (RH) is gradually increased. If the adhesive interactions are strong, a large amount of water will condense onto the substrate. The quantity of adsorbed water is given by the tradeoff between the energy cost of condensing water from an unsaturated vapor and the adhesive energy gain of having a water film at the surface \cite{bonn2}. We therefore study the adsorption of water films onto collections of graphene flakes with a large surface area; this allows us to establish the wetting properties of water on graphene alone. The presence of a very large surface area also minimizes the effects of surface contamination and possible defects in the graphene. The flakes are of different thicknesses, so we can measure the adsorption properties as a function of the number of graphene layers without relying on a non-graphene solid substrate.

\section*{Experiments and Results}
Graphene nanopowders (Graphene Supermarket, NY, USA) of three different flake thicknesses are used. We obtained the number of graphene layers (N) present in the nanopowder sample from Raman spectroscopy data (details in Supplementary Information) following the relation \cite{huiwang} $\omega_G$=1581.6+$\frac{11}{(1+N)^1.6}$, where $\omega_G$ is the wavenumber corresponding to the G-peak. The calculation yields N$\sim$1, N$\sim$ 4-5 and N$\sim$25, respectively, for the three nanopowder samples. A single graphene layer is $\sim$ 0.35 nm thick \cite{novoselov}. We will refer to the different thickness samples below by indicating the number of layers. The specific surface areas (measured by Brunauer-Emmett-Teller or BET method by the supplier) of N$\sim$1, N$\sim$ 4-5 and N$\sim$25 thick graphene nanopowder samples are 500 m$^2$/g, 100 m$^2$/g and 20 m$^2$/g, respectively.\\
\begin{figure}[h!]
\centering
\includegraphics[width=\textwidth]{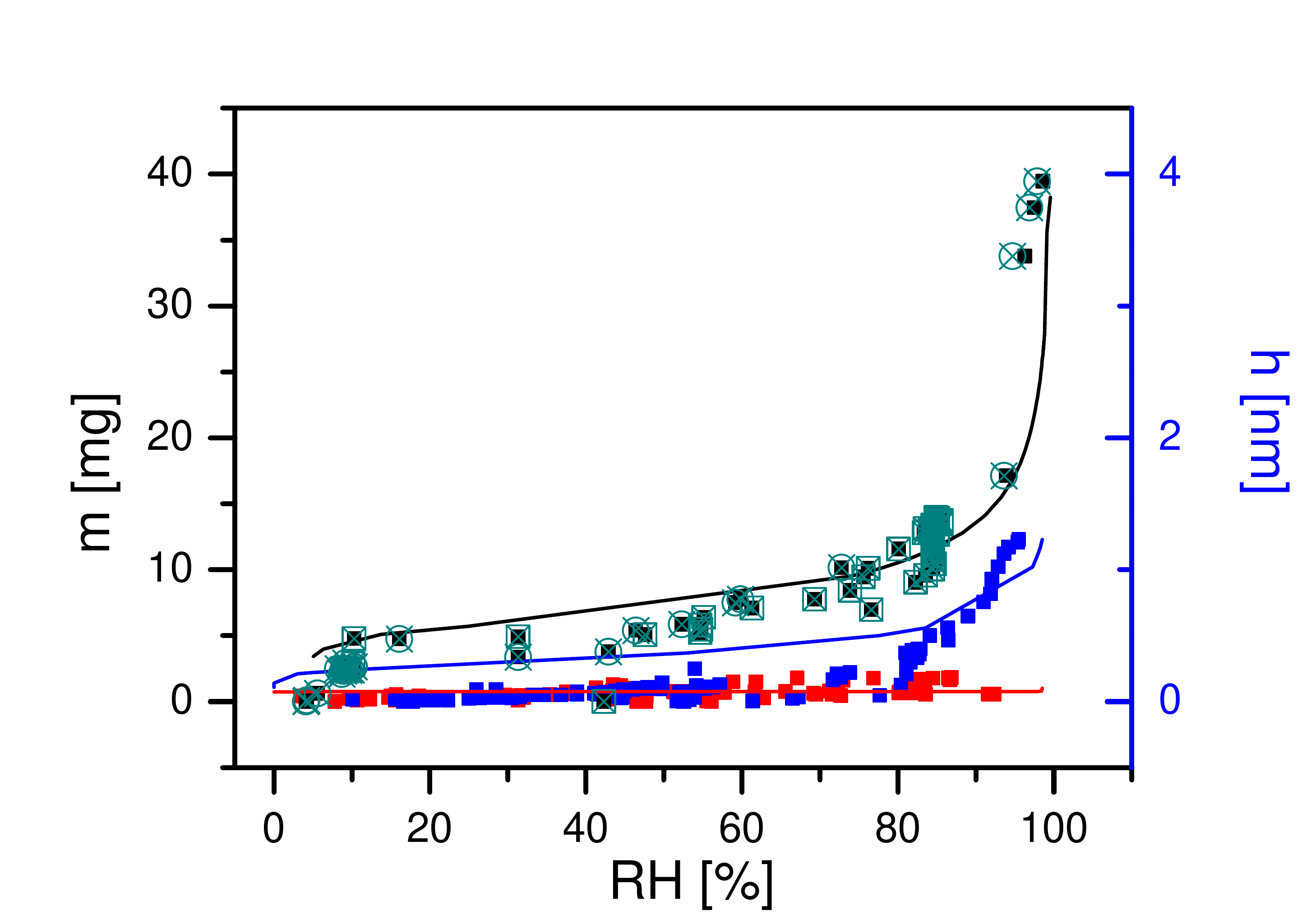}
\caption{Adsorbed water w.r.t RH for graphene nanopowder of various flake thicknesses. The relative thickness of these samples provided by the manufacturer (1:5:31) agrees rather well with our number of graphene layers calculation from Raman data (1:5:25). The various shape of symbols (square, circle and triangle) signify separate experiments. Continuous lines depict predicted moisture adsorption for graphene nanopowders based on various flake thickness values as provided by the manufacturer: 1 layer (red), 4-5 layers (blue) and 25 layers (black). In addition, the adsorbed film thickness $h$ in case of 50 nm thick graphene nanopowder is also shown with the crossed open symbols.}\label{fig1}
\end{figure}
The amount of water adsorbed by graphene nanopowder is measured on an automated balance with a precision of 0.0001 g placed in a controlled climatic chamber\cite{desarnaud} under nitrogen atmosphere to avoid any contamination. Figure \ref{fig1} shows the mass of water $m$ adsorbed by graphene which clearly varies with the number of graphene layers in the experiment as the relative humidity (RH) is changed. The layer of nanopowder with the thinnest flake size (N$\sim$1) does not adsorb any water even at the highest RH. In contrast, the nanopowder consisting of 4-5 layers of graphene adsorbs considerable amount of water, and the 25 layer thick flakes even more. In our experiments, we measure the adsorbed mass of water $m$; knowing the specific surface area $a$ and the amount of graphene nanopowder $m_0$, we can deduce the layer thickness of the film $h=\frac{m}{m_0~a~\rho}$, with $\rho$ the density of water (Fig.\ref{fig1}). We verified that the moisture adsorption trends show no hysteresis by comparing the results for increasing RH (from 5\% to 90\%) with decreasing RH (from 90\% to 5\%); this excludes capillary condensation between the graphene nanopowder flakes \cite{bonn1,bonn2} or at kinks and defects in the graphene. The Raman spectra show that in these experiments we do not probe pristine graphene layers. The flakes have quite some edges and kinks and bends, all of which will show in the Raman spectra. However, as long as the surface area of the graphene is significantly smaller than that of the edges and kinks (as is the case here), one can still reliably perform the water sorption experiments. With a density of $\sim$2 g/cm$^3$ and a typical flake size of 10 $\mu$m x10 $\mu$m x 1 nm (supplementary materials) one can calculate how much water molecules constitute a monolayer on the surface and on the edges. Taking 2 $\AA$ x 2 $\AA$ as the typical size of an adsorbed water molecule, one finds $\approx$ 10$^{22}$ molecules on the surface and $\approx$ 2.10$^{18}$ molecules on the edges. Hence the edge effects are negligible since they are 4 orders of magnitude smaller. Our observation that the water sorption is reversible and shows no hysteresis also reveals that there is no capillary condensation between the flakes, or at the edges.

\section*{Van der Waals forces}
In order to understand the mass adsorption characteristics, we start with the reasonable assumption that the adhesive interaction between the water and the graphene is due to the van der Waals forces. These can be calculated and compared with our adsorption experiments. Our system consists of a few molecular layers thick graphene (Fig.2a); in this case the adhesion force per unit area can be written as \cite{israel}:
\begin{figure}[h!]
\centering
\includegraphics[width=\textwidth]{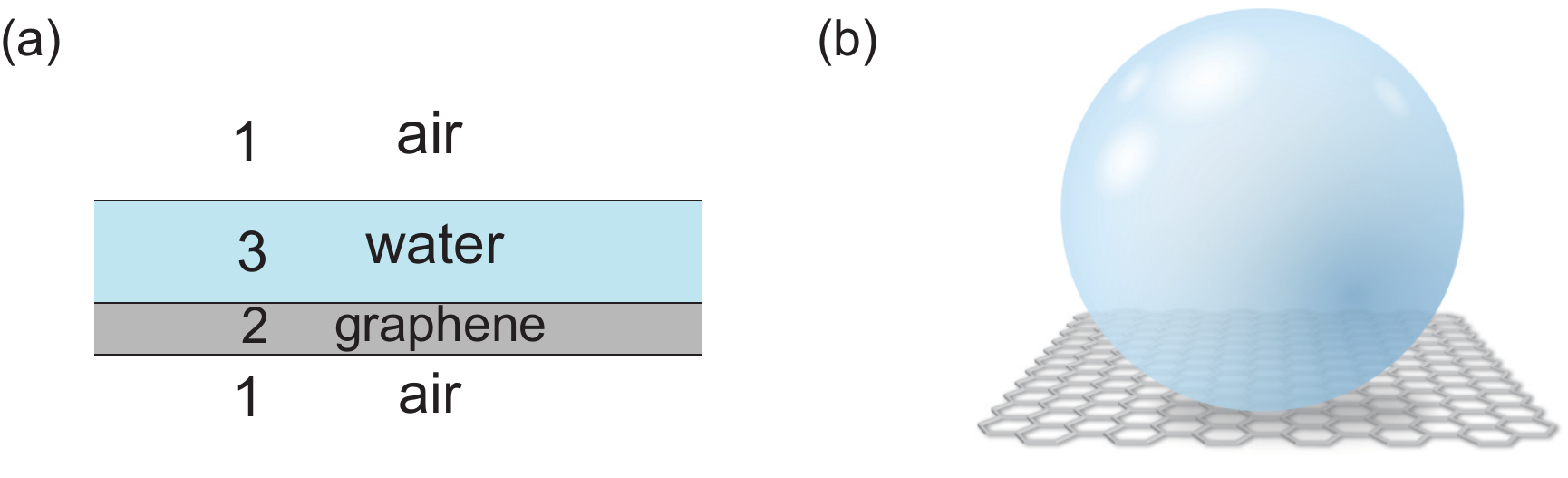}
\caption{(a) Schematic of graphene wetting represented as an adsorbed water film of thickness $h$. (b) Schematic of graphene's wetting transparency in air: the van der Waals interactions between graphene and water molecules are so much smaller than those between the water molecules themselves that the equilibrium contact angle is $\approx$ 180$^o$.}\label{fig2}
\end{figure}
\begin{equation}
\Pi_{vdw}(h)=-\frac{1}{6\pi}\frac{A_{132}}{h^3}-\frac{1}{6\pi}\frac{A_{12/31}}{(h+t)^3}
\end{equation}
where, A$_{132}$ is the Hamaker constant with interactions between the interfaces 12 (air-graphene) and 32 (water-graphene) in consideration, A$_{12/31}$ is that for interfaces 12 (air-graphene) and 31 (air-water), $t$ is the flake thickness of graphene nanopowder and $h$, the adsorbed water film thickness. The equilibrium film thickness $h_{eq}$ follows from the equilibrium between the adhesion forces and the free energy cost of condensing the water from the unsaturated vapor \cite{israel,bonn1}. It follows that:
\begin{equation}
\Pi(h_{eq})=\Pi_{vdw}+ \frac{\Delta \mu}{v_{w, molar}}=0
\end{equation}
where, $\Delta \mu$=RT $\ln (\frac{p}{p_{sat}})$=RT $\ln (\frac{RH}{100})$ is the chemical potential difference between the experimental and saturated relative humidity,  and $v_{w, molar}$ is the molar volume of water at room temperature.\\

The Hamaker constant A$_{12/31}$ can be can be calculated by noting that A$_{12/31}$=$\sqrt{A_{121}A_{313}}$. Using DLP theory \cite{israel}, the Hamaker constants have been calculated from the known refractive indices and the dielectric constants. The details of the calculation are provided in the Supplementary Information. Since we already know the adsorbed aqueous film thickness $h$ from the adsorbed mass $m$, Eq.1 and 2 can be solved for $\Delta \mu$.  The trend of adsorbed mass as a function of RH agrees quite well with the experiments (Fig.\ref{fig1}). The overall agreement is quite remarkable, given the fact that there are no adjustable parameters in the calculation. The slight overestimation, in the theory, of the water film thickness for low RH is likely to be due to the fact that the
the calculation of the Hamaker constants is not straightforward, since one of our phases is water with a high dielectric constant and many optical absorption bands \cite{bonn3}. \\

These results show that for the molecularly thin graphene, the mass of adsorbed water and hence the equilibrium film thickness is indistinguishable from zero, even \textit{at} coexistence i.e., at 100$\%$ relative humidity. In agreement with the experimental and calculated water adsorption data, the van der Waals adhesive interactions between a water drop and a single graphene layer are extremely weak, leading to a contact angle $\sim$180$^o$. The contact angle $\theta$ of the drop on the solid phase can be calculated directly from the disjoining pressure $\Pi(h)$;  the spreading parameter $S$ can be written as:
\begin{align}
S=\int_{h_{eq}}^{\infty}\Pi(h)dh
\end{align}
i.e., by calculating the work necessary against the van der Waals forces to go from a film of thickness h$_{eq}$ to an infinitely thick film\cite{bertrand}. Young's equation, then gives the contact angle as:
\begin{align}
\cos \theta&=1+\frac{S}{\gamma_{lv}}
% \sim 1-\frac{15*10^{-21} \textrm{m}^2}{h^2}
\end{align}
The calculation then leads to $\theta$=179.3$^o$, 163.2$^o$ and 139.7$^o$ ($\pm$2$^o$), for the three graphene flake thicknesses of N$\sim$1, N$\sim$ 4-5 and N$\sim$25; respectively: Indeed the molecularly thin graphene is superhydrophobic.\\\\
\begin{figure}[h!]
\centering
\includegraphics[width=\textwidth]{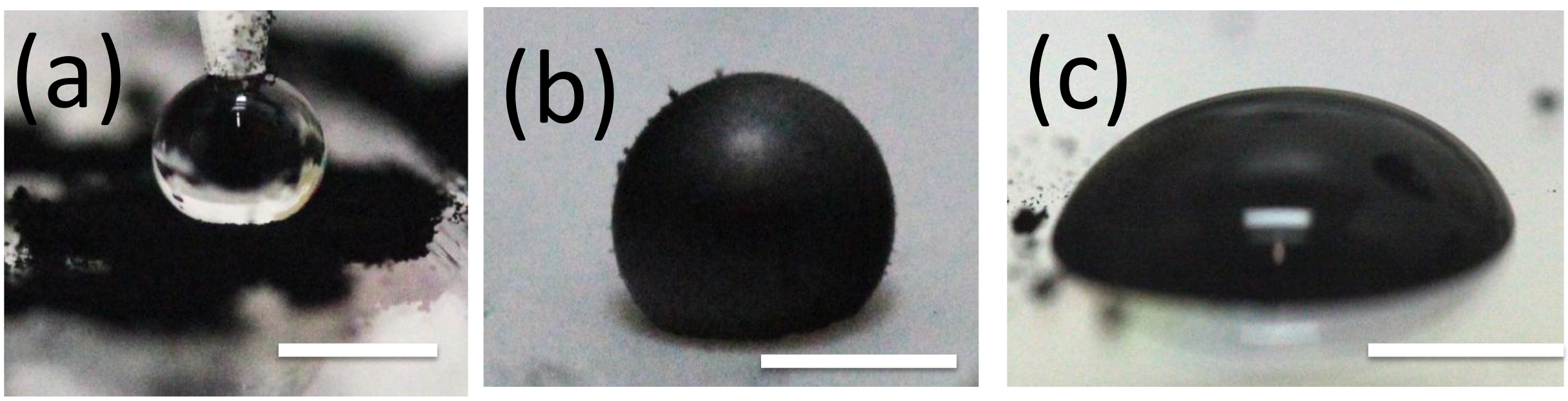}
\caption{Graphene powders of different flake thicknesses in contact with a water drop; (a) 1 layer grapheme (b) 4-5 layer graphene and (c) 25 layer graphene. The scale bars in A, B and C each represents a length of 2 mm. In (a) there is nothing on the surface, in (b) there is graphene on the liquid vapour interface in (c) there is graphene in the suspension but not at the surface.}\label{fig3}
\end{figure}
\section*{Test for Liquid Marbles}
An independent test of the observation that the contact angle of water on molecularly thin graphene is indistinguishable from 180$^o$ can be obtained by attempting to adsorb the graphene flakes onto a water drop. If the adhesive interaction is strong enough, the flakes will adhere to the liquid drop coating the whole surface in the process. This phenomenon is well known for hydrophobic particles adsorbing onto water drops; this leads to the formation of `liquid marbles', or pearls, which are liquid drops encapsulated by a hydrophobic powder. This results in a powder-coated drop that rolls very easily over any solid substrate and the  properties of such drops have been investigated in detail \cite{quere, supakar}. Theoretically, it is firmly established that particles of a material with a contact angle smaller than 180$^o$ should adsorb onto the surface of a liquid drop to form liquid marbles since the system gains the adhesion energy between the powder and the drop \cite{mchale}. In our `liquid graphene marble' experiments, the different graphene nanopowders are placed onto flat polystyrene substrates (experimental details in Supplementary Information). Putting a water droplet (volume 5 $\mu$l) onto the powder indeed leads to the spontaneous formation of a liquid marble for the intermediate thickness (4-5 layers) flakes (Fig.3b). As expected, the resulting drop has a very large contact angle and rolls easily over the surface. For the thickest flakes (25 layers), the powder covers the surface but a small part of the powder even intrudes into the water: the powder is no longer sufficiently hydrophobic to form perfect marbles, and a drop with a finite contact angle ensues, with the graphene being inside the drop rather than at its surface (Fig.3c). On the contrary, the molecularly thin flakes \textit{are not even adsorbed onto the surface} implying that there is no measurable adhesive interaction between the graphene and the water (Fig.3a). All these observations are in line with the mass measurements above, and confirm that the contact angle of water on molecularly thin graphene is $\sim$ 180$^o$.\\\\
These findings rationalize previous observations on the wetting properties and wetting transparency of graphene. The confusion in the literature is likely to be due to the effects of the substrate that significantly contribute to the adhesive interactions with the water drop. Our results are independent of a specific substrate. We treat graphene nanopowder as a porous medium, where the amount of water adsorbed follows a BET type of adsorption isotherm and thus only depends on the flake thickness of graphene. The wetting transparency of molecularly thin graphene is conclusively shown through our adsorption experiments, and is further supported by the calculation of van der Waals type of interactions in our system. If the substrate is air and graphene is wetting transparent, the contact angle is $180^o$, since the contact angle of a water drop in air is $180^o$. This is what we find. If the substrate is different, the van der Waals interaction of the liquid with the substrate is dominant over the graphene contribution, simply because the number of graphene atoms scales with the system size squared (2 dimensions) whereas the number of substrate atoms scales with the system size cubed (3 dimensions). It is for this reason that graphene is wetting transparent. The contact angle is always a tradeoff between the cohesion of liquid molecules in the drop that favor a $180^o$ angle (because a sphere has the largest volume to surface ratio), and the adhesion of the liquid on the surface that favors a smaller contact angle to enlarge the adhesive contact area. The adhesion is larger for thicker graphene since more carbon atoms contribute to the adhesive interaction with the surface. We thus close the ongoing debate on wetting transparency of graphene and thereby open the way for improved use of graphene in contact with water, such as coating, catalytic or electrochemical systems.

\bibliographystyle{abbrv}
\providecommand{\noopsort}[1]{}\providecommand{\singleletter}[1]{#1}%

\newpage

\chapter{\Large{Supplementary Information:Wetting of Water on Graphene}}

\section*{Characterization of graphene nanopowder}

In order to investigate the thickness and other properties of the graphene nanopowder flakes, various characterization methods such as a Nova-Nano high resolution SEM, transmission electron microscopy (TEM), energy dispersive X-ray spectroscopy (EDS) and electron energy loss spectroscopy (EELS) have been used. TEM imaging is done using an aberration-corrected, monochromated FEI Titan 80-300 microscope. In order to reduce the knock-on damage on graphene to a minimum, the microscope is operated at 60 kV. The dosage for TEM as well as the EDS and EELS is kept constant at 300 kV.\\\\\\

\begin{figure}[h!]
\centering
\includegraphics[width=\textwidth]{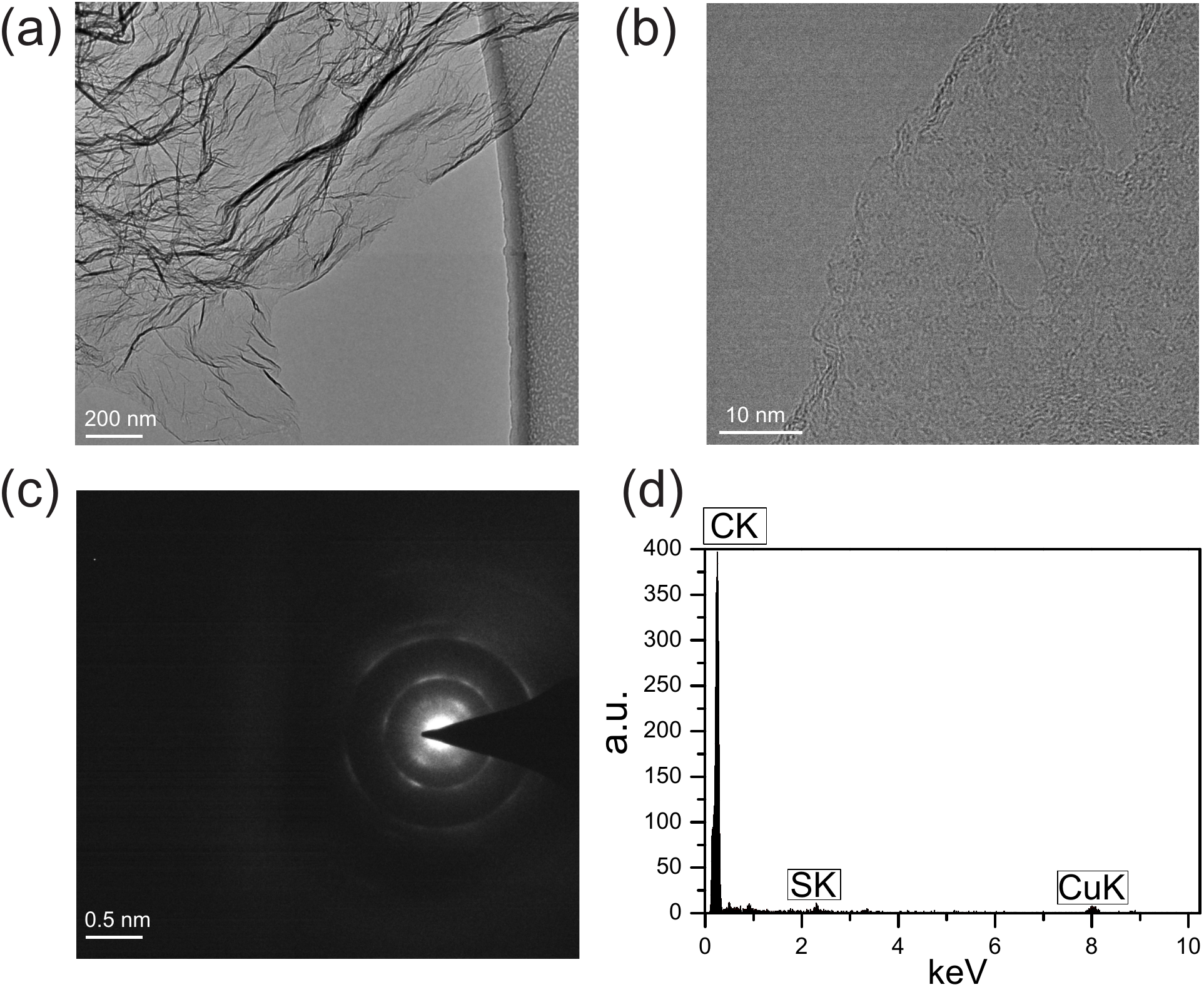}
\caption*{Figure S1: TEM images (a) and (b) for graphene nanopowder with 1 layer flake thickness qualitatively show that these flakes are thinnest among the samples. Selected area diffraction (c) shows the crystallinity of the sample and that the film is very thin. EDS (d) shows the purity of the sample.}\label{figS1}
\end{figure}

\begin{figure}[h!]
\centering
\includegraphics[width=\textwidth]{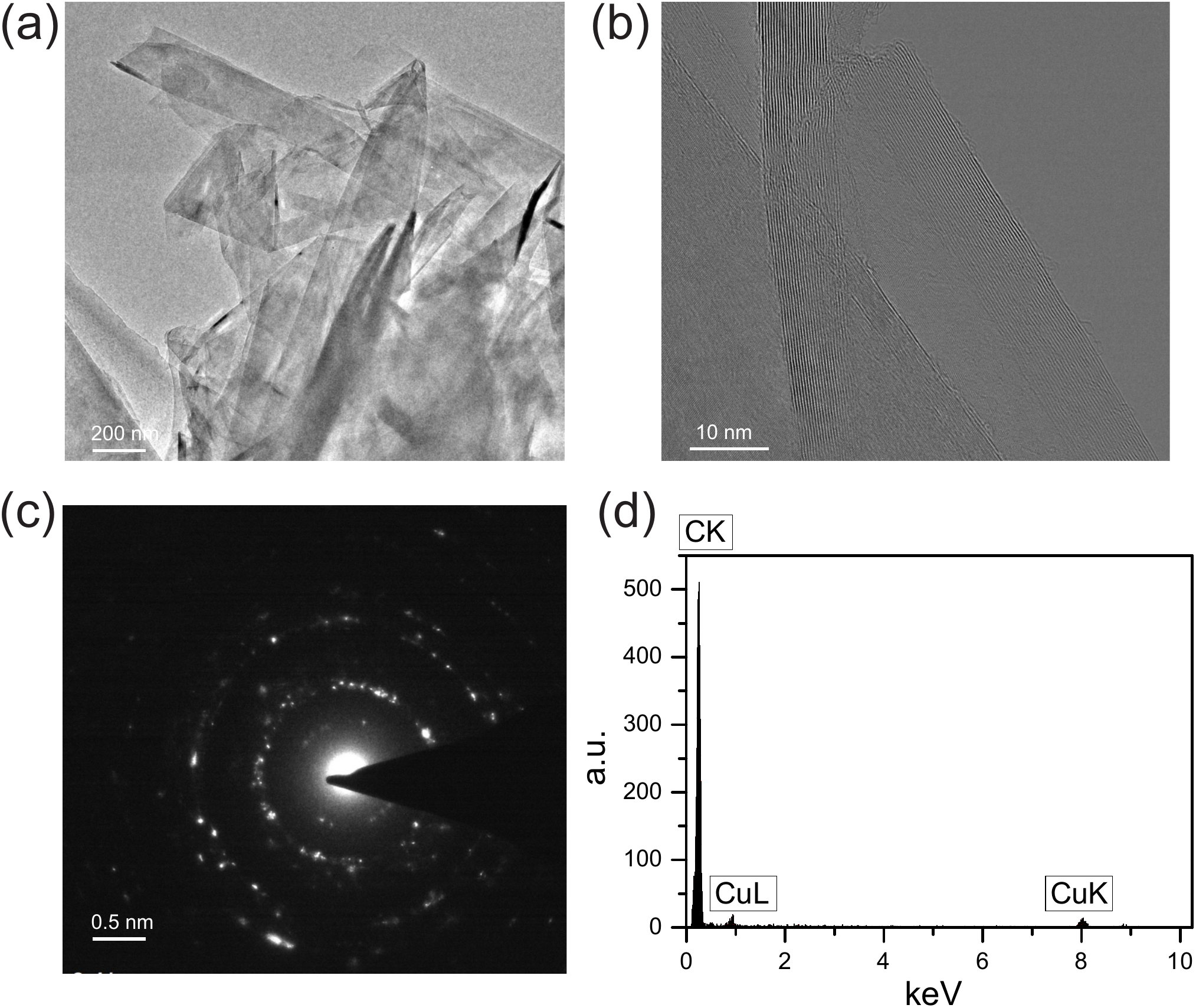}
\caption*{Figure S2: TEM images (a) and (b) for graphene nanopowder with intermediate flake thickness qualitatively show that these flakes are not the thinnest among the samples with onion-like flake structure. Selected area diffraction (c) shows the crystallinity of the sample and that the film is thicker (more crystalline volume) than depicted in Fig. S1. EDS (d) shows the purity of the sample.}\label{figS2}
\end{figure}

\begin{figure}[h!]
\centering
\includegraphics[width=\textwidth]{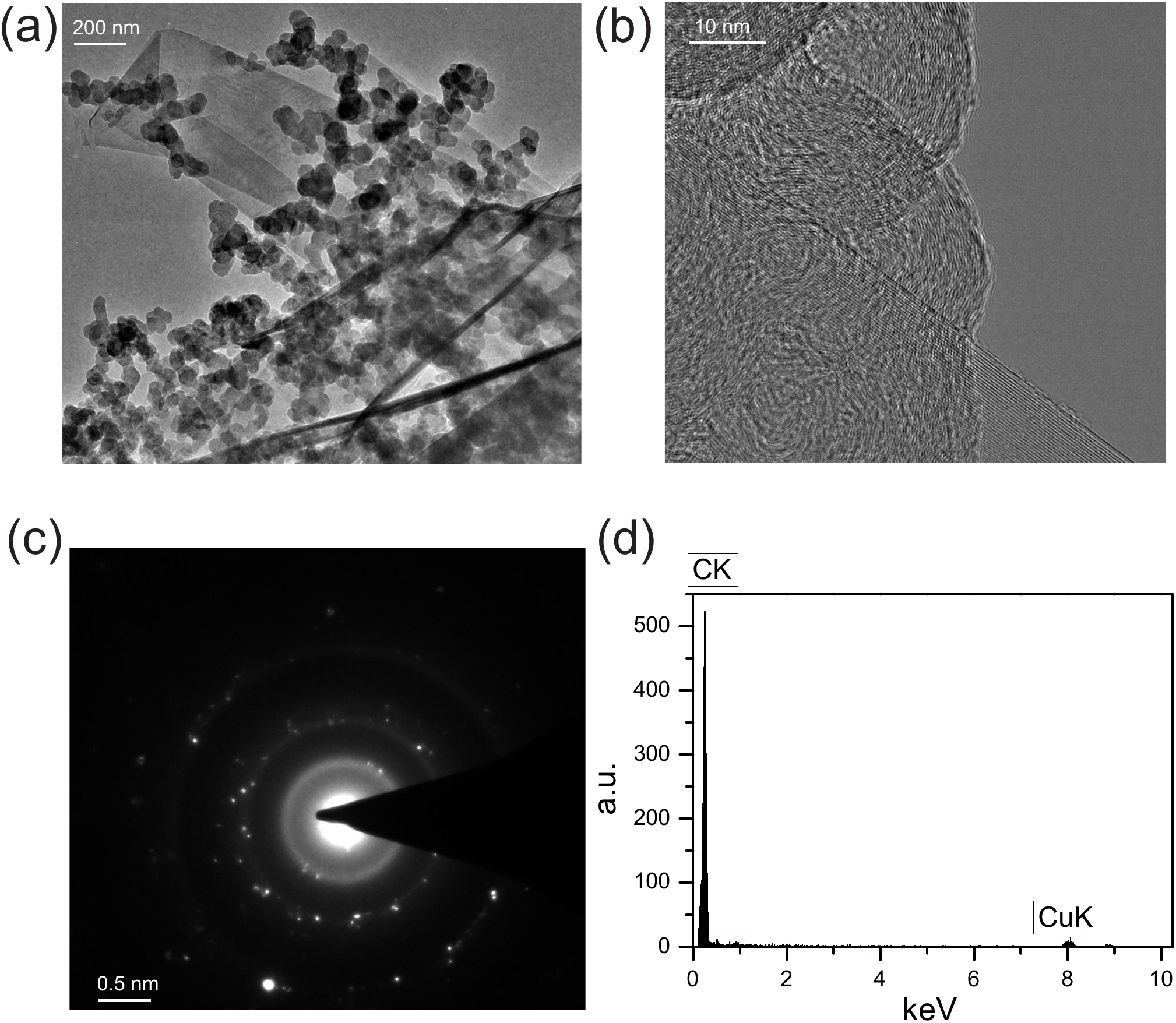}
\caption*{Figure S3: TEM images (a) and (b) for graphene nanopowder with $\sim$25 layer flake thickness qualitatively show that these flakes (platelets) are significantly different than the other samples. Selected area diffraction (c) shows the crystallinity of the sample and that the film is the thickest. EDS (d) shows the purity of the sample.}\label{figS3}
\end{figure}

\begin{figure}[h!]
\centering
\includegraphics[width=\textwidth]{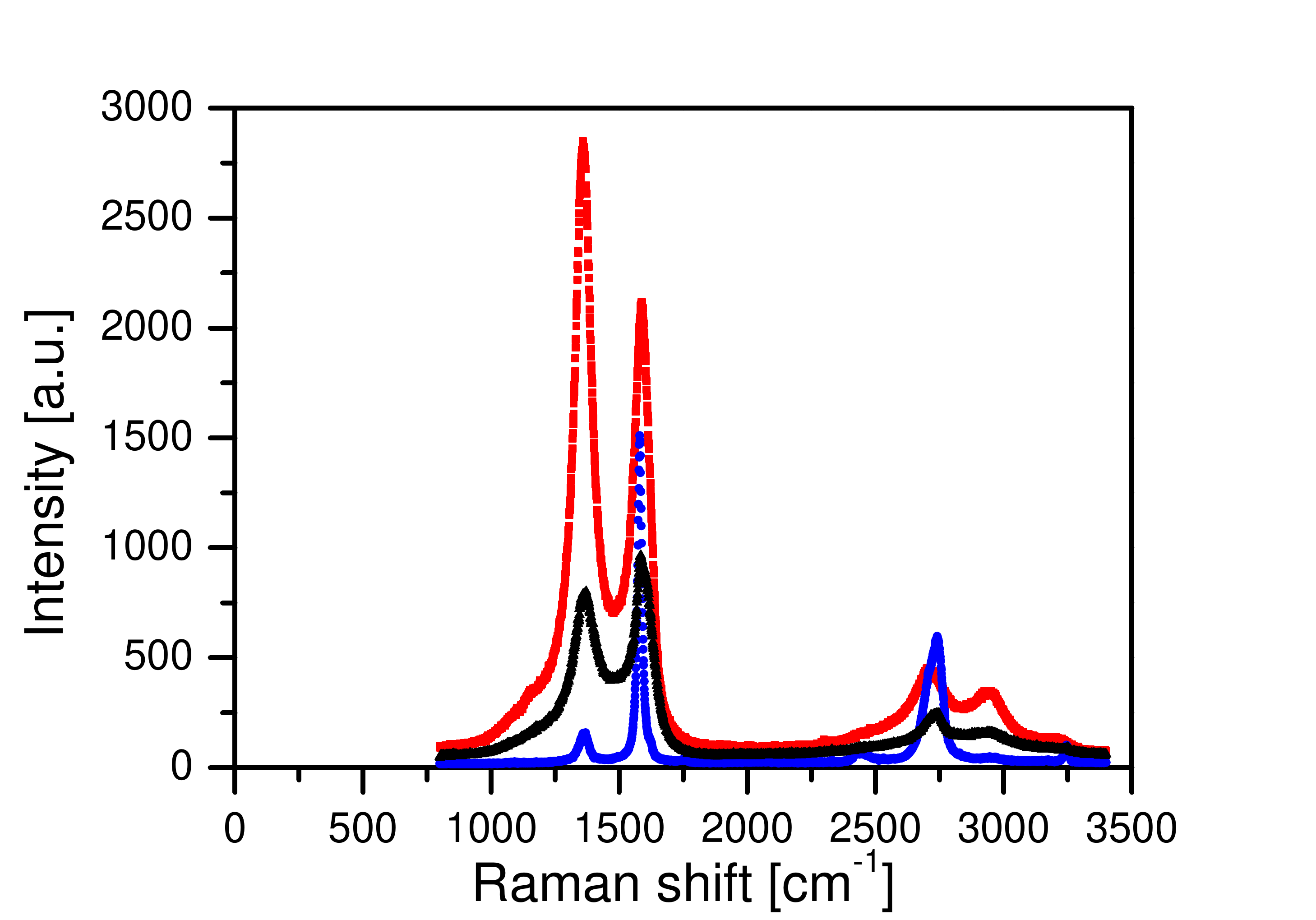}
\caption*{Figure S4: Raman spectra for the three graphene nanopowder samples: 1 layer (red), 4-5 layers (blue) and 25 layers (black).}\label{figS4}
\end{figure}
The properties of the different samples as given by the manufacturer are the following:\\\\
For the thinnest sample\\\\
Average Particle (lateral) size: $\approx$ 10 $\mu$ms.\\
Average X \& Y Dimensions:  $\approx$ 10 $\mu$ms.\\
Specific Surface Area : 400-800 m$^2$/g\\
True Density:$\leq$2.20 g/cm$^3$\\
Solids (Percent): $\geq 97.90 \%$\\
Carbon Content (percent of carbon): $\geq$ 95.00 \%\\
Oxygen Content (percent of oxygen): $\leq$ 2.50 \%\\\\\\
For the intermediate thickness\\\\
Specific surface area:100 m$^2$/g\\
Purity: $99.9 \%$\\
Average flake thickness: 8 nm\\
Average Particle (lateral) size: $\approx$ 550 nm (150-3000) nm.\\\\
And for the largest thickness:\\\\
Specific surface area $\leq$ 20 m$^2$/g\\
Purity: $98.5 \%$ \\
Average flake thickness: 50 nm\\
Particle (lateral) size: $\approx$ 3-7 $\mu$ms\\.

For the Raman spectra, it is worthwhile noting that Large isolated defect-free single-layer graphene shows two sharp intense bands at 1580 cm$^{-1}$ (G) and ca. 2690 cm$^{-1}$ (2D, excitation wavelength dependent position), with I(2D)/I(G)$\sim$2. The smaller I(2D)/I(G), the larger number of layers in a first approximation bc of layer-layer interactions. More layers (i.e. more interactions) also causes a red shift of G and a splitting, widening and upshift of 2D. The D band at 1350 cm$^{-1}$ is generally symmetry forbidden in Raman spectroscopy of pristine graphene. Furthermore, with increasing number of defects, i.e. increasing electron-phonon scattering, other ‘disorder’ peaks appear: 1350 cm$^{-1}$ (D), 1620 cm$^{-1}$ (D’), 2940 cm$^{-1}$ (D+G). People use the I(D)/I(G) ratio to estimate the characterize the level of disorder. Typically, I(D)/I(G) increases with increasing defect density in a ‘low defect density’ regime (more elastic scattering bc of more defects), up to a certain point where the material becomes too amorphous and scattering is attenuated and (I(D)/I(G) drops again.\\\\
The Raman spectra therefore clearly show that we do not probe pristine graphene layers. The flakes have quite some edges and kinks and bends - all of which show in the Raman spectra. However as long as the surface area of the graphene is significantly smaller than that of the edges and kinks (as is the case here) one can still reliably perform the water sorption experiments. Our observation that the water sorption is reversible and shows no hysteresis also reveals that there is no capillary condensation between the flakes, or at the edges.

\section*{Water Adsorption Experiments}

\begin{figure}[h!]
\centering
\includegraphics[width=\textwidth]{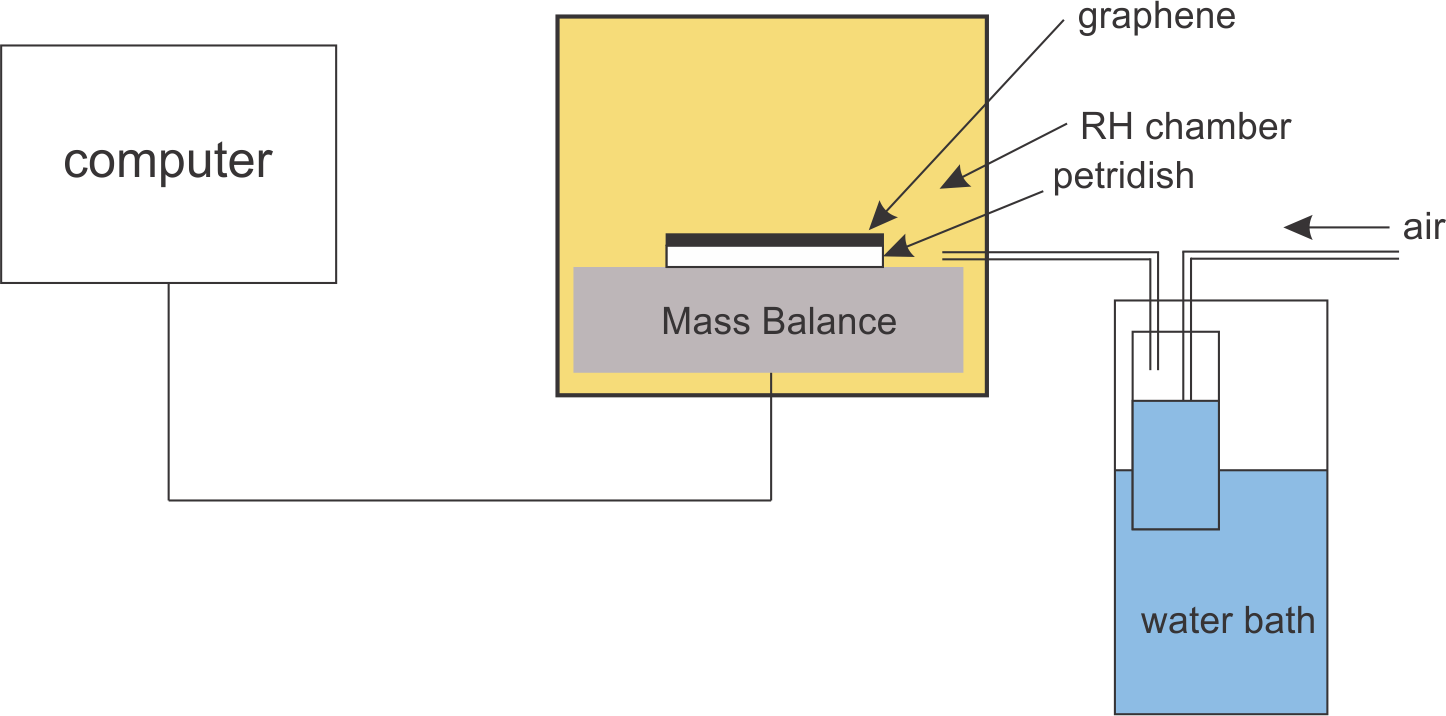}
\caption*{Figure S5: Schematic of the water adsorption experiments.}\label{fig4}
\end{figure}

The water adsorption experiments (Fig.S5) are performed using a precision mass balance (New Classic Balances MS-S/MS-L, Mettler Toledo, USA) with a precision of four decimal points of 1 g.  The mass balance itself is housed in a protected environment (nitrogen atmosphere) of a climatic chamber in order to control the relative humidity (RH) of the environment surrounding the samples.. The sample under study is also prepared in a protected environment. The graphene nanopowder is placed on an open polystyrene or glass petridish (of diameter 5.5 cm) and subsequently transferred inside the mass balance, taking extreme precautions so that the outside relative humidity (RH) and contaminants do not affect the sample. The initial mass (m$_0$) is not the same from sample to sample, instead the surface areas of the samples, i.e., [specific surface area (BET) $\times$ initial mass (m$_0$)]  are kept constant during the experiments.The mass of the sample is recorded every 10 minutes during the experiment using Mettler Toledo Balancelink software. The RH at the climatic chamber is fixed by controlling $p_w$, the partial water vapor pressure. A water bath thermostat (Haake A 25) is used to let air flow through water at a specific temperature T$_1$, which then proceeds to the climatic chamber at the room temperature T$_2$=21$^o$C. Hence, partial vapor pressure at temperature T$_2$ equals the saturated vapor pressure at temperature T$_1$, and the RH is given by: RH=$\frac{p_{ws}(T_1)}{p_{ws}(T_12}$.100.\cite{desarnaud} In our experiments, RH has been changed from 5\% to 90\% in steps of 10\%, where the temperature T$_1$ is varied between 5$^o$ and 25$^o$ in order to achieve the desired RH. Once a new temperature is set in the water bath, $\sim$ 45 minutes is allowed for reaching the desired RH, and subsequently, we wait for 3 hours for the RH and mass to stay at the equilibrium value. The RH at the mass balance chamber is measured using a hygrometer ($Testo$ 625) and the values are also recorded at regular time intervals.\\

\section*{Calculation of van der Waals forces}
\begin{figure}[h!]
\centering
\includegraphics[width=0.3\textwidth]{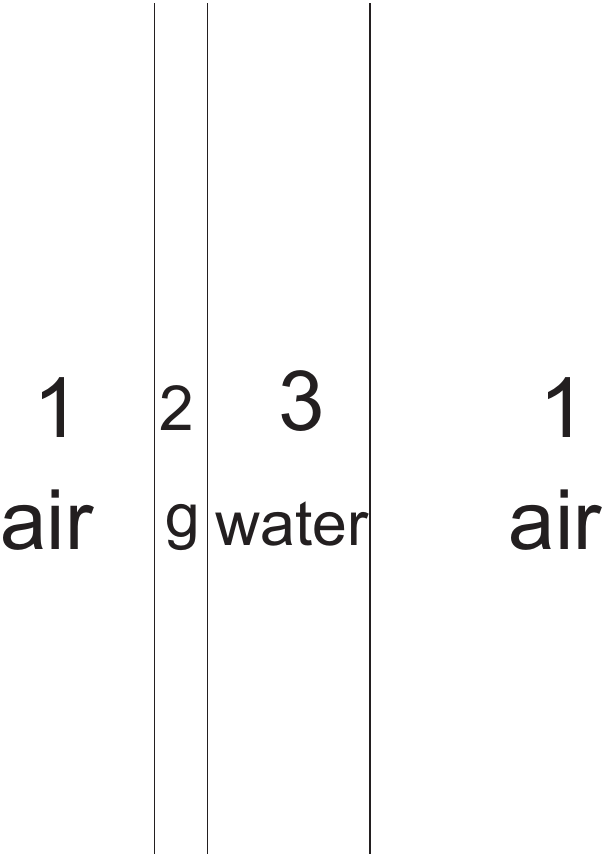}
\caption*{Figure S6: Schematic for calculation of van der Waals contribution.}\label{fig4}
\end{figure}
We consider two semi-infinite media (air, denoted by phase 1), interacting to each other through a layer of water (medium 3, thickness $h$), when one of the air phases has an adsorbed layer of graphene (medium 2) of thickness $t$. Then, following Israelachvili \cite{israel}, Bonn \textit{et al.} \cite{bonn1}, Bonn \& Ross \cite{bonn2}, the disjoining pressure $\Pi$ is given by:
\begin{equation}
\Pi_{vdW}(h)=-\frac{1}{6\pi}\frac{A_{132}}{h^3}-\frac{1}{6\pi}\frac{A_{12/31}}{(h+t)^3}=-\frac{1}{6\pi}\frac{A_{132}}{h^3}-\frac{1}{6\pi}\frac{\sqrt{A_{313}A_{121}}}{(h+t)^3}]
\end{equation}
where, from Lifshitz Theory,
\begin{align}
A_{132}&=A_{\textrm{air(1)graphene(2)interacting across water(3)}} \nonumber \\
&=\frac{3}{4}k_BT(\frac{\epsilon_a-\epsilon_w}{\epsilon_a+\epsilon_w})(\frac{\epsilon_g-\epsilon_w}{\epsilon_g+\epsilon_w})+\frac{3h_p\gamma_e}{8\sqrt{2}}\frac{(\eta_a^2-\eta_w^2)(\eta_g^2-\eta_w^2)}{(\eta_a^2+\eta_w^2)^{1/2}(\eta_g^2+\eta_w^2)^{1/2}[(\eta_a^2+\eta_w^2)^{1/2}+(\eta_g^2+\eta_w^2)^{1/2}]} \nonumber \\ 
A_{313}&=A_{\textrm{water(3) air(1)water(3)}} \nonumber \\
&=\frac{3}{4}k_BT(\frac{\epsilon_a-\epsilon_w}{\epsilon_a+\epsilon_w})^2+\frac{3h_p\gamma_e}{8\sqrt{2}}\frac{(\eta_a^2-\eta_w^2)^2}{(\eta_a^2+\eta_w^2)^{1/2}(\eta_a^2+\eta_w^2)^{1/2}[(\eta_a^2+\eta_w^2)^{1/2}+(\eta_a^2+\eta_w^2)^{1/2}]} \nonumber \\ 
A_{121}&=A_{\textrm{air(1) grap(2) air(1)}} \nonumber \\
&=\frac{3}{4}k_BT(\frac{\epsilon_a-\epsilon_g}{\epsilon_a+\epsilon_g})^2+\frac{3h_p\gamma_e}{8\sqrt{2}}\frac{(\eta_a^2-\eta_g^2)^2}{(\eta_a^2+\eta_g^2)^{1/2}(\eta_a^2+\eta_g^2)^{1/2}[(\eta_a^2+\eta_g^2)^{1/2}+(\eta_a^2+\eta_g^2)^{1/2}]}
\end{align}
where $\epsilon$ is the dielectric constant of the medium (abbreviated in the subscript), $\eta$ is the refractive index, $\gamma_e$ is the electron adsorption frequency and $h_p$ is the Planck's constant. $\epsilon_g$=7.5 and $\eta_g$=2.5 are used in our calculation.\\
We can write the total interaction potential of the system as $\Phi(h)$ as:
\begin{equation}
\Phi(h)=\Phi_{vdw}+\Delta \mu \frac{1}{v_{w, molar}} h
\end{equation}
where, $\Delta \mu$=RT $\ln (\frac{p}{p_{sat}})$=RT $\ln (\frac{RH}{100})$\cite{bonn1} is the change in the chemical potential brought forward by the change of RH in the surrounding and $v_{w, molar}$ is the molar volume of water at room temperature. Since at equilibrium, $\frac{d\Phi(h)}{dh}=0$:
\begin{align}
\frac{d\Phi_{vdw}(h)}{dh}+\frac{\Delta \mu}{v_{w, molar}}&=0 \nonumber \\
\textrm{or,} -\frac{1}{6\pi}\frac{A_{132}}{h^3}-\frac{1}{6\pi}\frac{\sqrt{A_{313}A_{121}}}{(h+t)^3}+\frac{\Delta \mu}{v_{w, molar}}&=0
\end{align}
If $m$ is the mass of adsorbed water on graphene, then: 
\begin{align}
m&=m_0.a.h.\rho \nonumber \textrm{[a in m$^2$/g, h in m and $\rho$ is 1 gm/ml]} \nonumber \\
\textrm{or,} m&=m_0.a.h*10^6 \textrm{[converting ml to m$^2$]} \nonumber \\
\textrm{or,} h&=\frac{m}{m_0.a*10^6}
\end{align}
where $m_0$ is the initial mass of the sample, $a$ is the specific surface area (BET area) with the thickness of the water layer ($h$) and the density $\rho$. Hence,
\begin{align}
-\frac{1}{6\pi}\frac{A_{132}}{h^3}-\frac{1}{6\pi}\frac{A_{12/31}}{(h+t)^3}+\frac{\Delta \mu}{{v_{w, molar}}}=0 \nonumber \\
-\frac{A_{132}}{h^3}-\frac{A_{12/31}}{(h+t)^3}+6 \pi \frac{RT}{v_{w, molar}}\ln (RH) &=0 \nonumber \\
-\frac{A_{132}(m_0.a*10^6)^3}{m^3}-\frac{\sqrt{A_{313}A_{121}}}{(\frac{m}{m_0.a*10^6}+t)^3}]+6 \pi \frac{RT}{v_{w, molar}}\ln (RH)&=0
\end{align}

\end{document}